# Capacity of the Aperture-Constrained AWGN Free-Space Communication Channel

Richard J. Barton, Member

*Abstract*— In this paper, we derive upper and lower bounds as well as a simple closed-form approximation for the capacity of the continuous-time, bandlimited, additive white Gaussian noise channel in a three-dimensional free-space electromagnetic propagation environment subject to constraints on the total effective antenna aperture area of the link and a total transmitter power constraint. We assume that the communication range is much larger than the radius of the sphere containing the antennas at both ends of the link, and we show that, in general, the capacity can only be achieved by transmitting multiple spatially-multiplexed data streams simultaneously over the channel. Furthermore, the lower bound on capacity can be approached asymptotically by transmitting the data streams between a pair of physically-realizable distributed antenna arrays at either end of the link. A consequence of this result is that, in general, communication at close to the maximum achievable data rate on a deep-space communication link can be achieved in practice if and only if the communication system utilizes spatial multiplexing over a distributed MIMO antenna array. Such an approach to deep-space communication does not appear to be envisioned currently by any of the international space agencies or any commercial space companies. A second consequence is that the capacity of a long-range free-space communication link, if properly utilized, grows asymptotically as a function of the square root of the received SNR rather than only logarithmically in the received SNR.

*Index Terms*— MIMO, deep-space communication, distributed antenna arrays, prolate spheroidal wave functions, Shannon capacity

## I. Introduction

The intent of this paper is to investigate the capacity of the free-space communication channel when available spatial diversity is incorporated into the channel model itself. In particular, we derive upper and lower bounds as well as a simple closed-form approximation for the capacity of the continuous-time, complex-valued, bandlimited, free-space, additive white Gaussian Noise (AWGN) channel for the scenario conceptually encountered in long-range deep-space communication applications. That is, information is assumed to be transmitted over a bandlimited radio channel in free space over a very large distance from an information source (the transmitter) to an information sink (the receiver) subject to constraints on both the total transmitter power and the total effective aperture areas of the transmitter and receiver antennas; however, the antennas at both the transmitter and the receiver may be distributed in an arbitrary fashion over spherical regions of space at both ends of the communication link subject only to the constraint that the radii of both regions is much smaller than the distance between them. We refer to this constraint simply as the *far-field assumption* because it allows us to make what amounts to a far-field approximation in the channel propagation model. In this scenario, it is reasonable to assume that the rate at which information can be distributed among all of the elements at each end of the link is much higher than the rate at which information can be distributed between the two spherical volumes and that the power required to distribute that information among the elements is much lower than the power required to distribute the information between the two volumes. In fact, we make the assumption that both the time and power required to distribute information among the elements of both the source and sink are insignificant to the determination of the maximum achievable rate at which information can be transmitted from the source to the sink.

The results presented in this paper justify and generalize the results presented in an earlier paper on deep-space MIMO communications [1]. In [1], the characteristics and benefits of MIMO for space application are discussed somewhat heuristically under the assumption that the multiplexing gain for a MIMO system in space scales with the number of antennas in exactly the same manner as that of a conventional terrestrial MIMO system operating in a rich scattering environment, with a simple constraint on total aperture area added to the equation. In the current paper, the information capacity of the free-space channel is studied in a much more mathematically rigorous context in order to both justify and generalize the heuristic results presented previously in [1].

Since the seminal papers on multiple-input, multiple-output (MIMO) additive Gaussian noise channels were published in the late nineties, there have been numerous studies on the capacity of the Gaussian MIMO channel under various assumptions regarding number of antennas, antenna configuration, and propagation environment. In many of these papers [2-6], capacity in a terrestrial fading environment is studied as a function of the number and configuration of the transmitter and receiver antennas, and the channels are generally modeled as random. The channels are characterized in terms of the correlation structure of the components of the channel matrix that models the coupling between transmitter and receiver antennas. Both the diversity gain and the multiplexing gain available in such environments depend critically on this correlation structure.

In several other papers [7-12], the focus is neither on the capacity of the channel, per se, nor on the number and configuration of the antennas, but rather on the modeling of communication channels in one-, two-, and three-dimensional electromagnetic propagation environments. In this case, the channels are not modeled as random, and the primary focus is often on identification of the mathematical structure of the operators that model propagation in the environment.

In yet another large class of papers, the focus is on the rate at which expected capacity scales as the number of nodes (antennas) in a communication network grows large. In this case, both the channel models and the antenna density and distribution are generally modeled as random. In some of these [13-15], the physical environment is not specifically considered, but of primary interest for the current work are the papers that specifically consider the impact of the physical propagation environment on the scaling of capacity in a MIMO context [16-17].

In the current paper, we seek to identify bounds on the capacity of the long-range, free-space AWGN channel, which is assumed fixed and non-random, without explicit reference to the number of antennas deployed or the distribution of those antennas. The result is derived from the properties of the operator that models free-space propagation over long distances rather than from the number and configuration of the antennas at each end of the link. At the same time, we demonstrate that the capacity on the channel can be approached asymptotically using physically realizable MIMO antenna arrays that sample the operator in discrete space. The effects of the density and distribution of the spatial sampling, and in particular numerical simulations of those effects, are beyond the scope of the current paper, but are clearly relevant and of interest for future studies.

Finally, since the work in this paper is possibly of most interest for deep space communication, it is worth noting that the impact of MIMO communication techniques using distributed antenna arrays in satellite communication has been considered in several different studies in the past. The majority of these are concerned primarily with distributed beam-forming in one form or another rather than more fundamental channel capacity questions. A particularly nice review of much of that work is presented in [18].

*A. Review of SISO and MIMO results*

To motivate the problem we are trying to solve as well as the rather abstract channel model that we adopt in this paper, we consider two simple examples. Once again, for both of these examples and throughout this paper, we assume that all channels are continuous-time, complex-valued, bandlimited, free-space, AWGN channels. As such, channel inputs, outputs, and noise processes are generally represented by functions of the form $y(t)$, $x(t)$, and $n(t)$, respectively (or vector-valued equivalents), but we will suppress the reference to time throughout the paper. This means that the simple vector notation **y** is shorthand for

$$\mathbf{y} = \mathbf{y}(t) = \left[ y_1(t), y_2(t), \ldots, y_n(t) \right]^T,$$

and that channel outputs will always be demodulated, sampled at the Nyquist rate, and arranged into arbitrarily long blocks of complex-valued symbols representing codewords for purposes of decoding.

For the first example, we assume that the transmitter has a single antenna with effective aperture area $A_T$ and the receiver has a single antenna with effective aperture area $A_R$. In this case, the maximum achievable data rate in bits/sec (b/s) on the channel is given by the *channel capacity*, which takes the form

$$C = B \log_2 \left( 1 + \frac{A_T A_R L P}{\lambda^2 d^2 B N_0} \right), \tag{1.1}$$

where $B$ is the bandwidth of the transmitted signal, $P$ is the transmitted power, $d$ is the range between the two antennas, $\lambda$ is the wavelength at the carrier frequency, $N_0$ is the power spectral density of the AWGN on the baseband equivalent (i.e., complex-valued) channel, and $L$ is a factor that represents the cumulative effect of additional unmodeled losses on the channel such as circuit losses, pointing losses, polarization losses, etc. See [19, (9.164)] for the general bandlimited AWGN channel capacity formula and [20, §5.5.2] for a discussion of the received power on bandlimited channels.

Equation (1.1) is generally expressed equivalently in terms of maximum achievable spectral efficiency (capacity per unit bandwidth) in bits per second per Hertz (b/s/Hz) as

$$\xi_1 = \frac{C}{B} = \log_2 \left( 1 + \frac{A_T A_R L P}{\lambda^2 d^2 B N_0} \right) = \log_2 \left( 1 + \gamma g \right), \tag{1.2}$$

where $g$ represents the *channel gain* and $\gamma$ represents the transmitted *signal-to-noise ratio* (SNR), which are given by

$$g = \frac{A_T A_R L}{\lambda^2 d^2} \text{ and } \gamma = \frac{P}{BN_0}, \qquad (1.3)$$

respectively. Note that $\gamma g$ represents the received SNR.

As a second example, consider the situation illustrated in Figure 1 below. In this case, the communication system has been *fractionated* into $M$ stations at each end of the link, where each station is equipped with a single transceiver with transmit power $P/M$, the stations at one end all have single antennas with aperture area $A_T/M$ and are distributed over the spherical volume $\mathcal{V}_T$, and the stations at the other end all have single antennas with aperture area $A_R/M$ and are distributed over the spherical volume $\mathcal{V}_R$. That is, the single-input, single-output (SISO) system architecture has been replaced by a distributed multiple-input, multiple-output (MIMO) architecture. Assuming that the unmodeled losses are the same on each of the point-to-point channels between stations at each end of the link, the channel gain on each individual point-to-point channel will be well approximated by $g/M^2$ and the transmitted SNR on each channel will be given by $\gamma/M$ [1]. The remaining behavior of the channel is then determined by the structure of the channel matrix **H**, which is given by

$$\mathbf{H} = \begin{bmatrix} h_{11} & h_{12} & \cdots & h_{1M} \\ h_{21} & h_{22} & \cdots & h_{2M} \\ \vdots & \vdots & \cdots & \vdots \\ h_{M1} & h_{M2} & \cdots & h_{MM} \end{bmatrix},$$

where $\{h_{ij}\}$, $i,j = 1,2,\ldots,M$, represent the complex-valued amplitude and phase couplings between receive antenna $i$ and transmit antenna $j$.

Whatever the actual structure of **H**, the maximum achievable spectral efficiency of the equivalent MIMO channel is then given by [2]

$$\xi_M(\mathbf{H}) = \log_2\left(\det\left[\mathbf{I} + \frac{\gamma g}{M^3}\mathbf{HH}^*\right]\right) = \sum_{i=1}^{M} \log_2\left(1 + \frac{\gamma g}{M^3}|v_i|^2\right), \qquad (1.4)$$

where $\mathbf{H}^*$ is the complex-conjugate transpose of the matrix **H** and $\{|v_1|^2 \geq |v_2|^2 \geq \cdots \geq |v_M|^2\}$ are the eigenvalues of $\mathbf{HH}^*$.

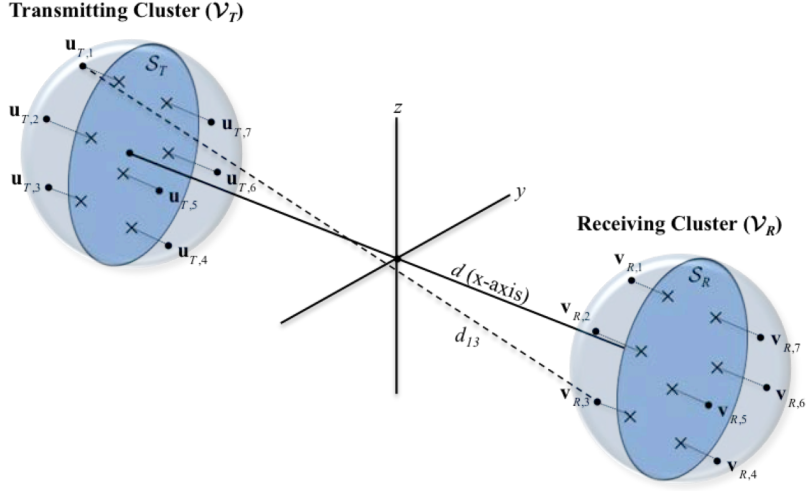

Figure 1. Fractioned MIMO Example with *M*=7.

Note that the spectral efficiency given by Equation (1.4) corresponds to channel capacity without channel side information at the transmitter. That is, if the vector channel model (at an arbitrary observation time) is represented by

$$\mathbf{y} = \sqrt{\frac{g}{M^2}} \mathbf{H} \mathbf{x} + \mathbf{n}, \qquad (1.5)$$

where $\mathbf{n} \sim \mathcal{N}(\mathbf{0}, N_0 \mathbf{I})$ (i.e., an iid vector-valued, continuous-time, bandlimited, complex AWGN noise process with power spectral density $N_0$), then (1.4) represents the maximum achievable spectral efficiency under the constraint $E\{\mathbf{x}\mathbf{x}^*\} = \frac{P}{M}\mathbf{I}$, which can be achieved using a codebook chosen from $\mathbf{x} \sim \mathcal{N}(\mathbf{0}, \frac{P}{M}\mathbf{I})$. If we assume that full channel side information is available at the transmitter, and we impose only a total transmitted power constraint, then the maximum achievable spectral efficiency is given by the so-called *waterfilling solution*, which takes the form

$$\xi_M^{wf}(\mathbf{H}) = \sum_{i=1}^{K} \log_2\left(1 + |v_i|^2 \left[\frac{\gamma g}{KM^2} + \frac{1}{K}\sum_{j=1}^{K}\frac{1}{|v_j|^2} - \frac{1}{|v_i|^2}\right]\right), \qquad (1.6)$$

where *K* is the largest integer such that

$$\frac{\gamma g}{KM^2} + \frac{1}{K}\sum_{j=1}^{K}\frac{1}{|v_j|^2} \geq \frac{1}{|v_i|^2}, \qquad (1.7)$$

for all $1 \leq i \leq K$. See [19, (9.165)-(9.168)] for the general formulae for Gaussian channel capacity with and without feedback and [21, §2.3.2] for a discussion of the effects of channel side information in the MIMO case.

### B. Goals and Organization

In this paper, we generalize the result given by Equations (1.6) and (1.7) to the case in which an antenna may be any two-dimensional surface (not necessarily connected) distributed arbitrarily in $\mathcal{V}_T$ or $\mathcal{V}_R$ that satisfies the required effective aperture area constraint in that region. Similarly, the power distribution over the radiating surface of the transmitter antenna may be any function that satisfies the total transmitted power constraint. For this very general expression of the long-range free-space communication channel, we develop the following results.

1. We derive upper and lower bounds on the maximum achievable spectral efficiency for this channel and show that, for values of the received SNR below a weak-signal threshold, these bounds are equal and equivalent to the well-known SISO result given by Equation (1.2).

2. For values above the weak-signal received SNR threshold, we also derive a simple approximation for the maximum achievable spectral efficiency, which shows that the capacity for the long-range free-space channel actually strictly exceeds the result given by Equation (1.2) and grows asymptotically as a function of the square root of the received SNR rather than logarithmically as in Equation (1.2).

3. Perhaps equally important from a practical point of view, we show that for values above the weak-signal threshold, the lower bound on the maximum achievable spectral efficiency can be approached arbitrarily closely using physically realizable distributed MIMO antenna arrays satisfying the antenna aperture constraints at each end of the link. This may have significance for the design of future high-data-rate deep-space communication systems.

The remainder of this paper is organized as follows. In Section II, we show that the far-field assumption allows us to study the capacity of the general free-space AWGN channel by restricting the distribution of the radiating surface of the antennas to a two-dimensional region of space around the source and the sink. The main results of the paper are then stated and proven in Section III in the form of Theorem 1 and its Corollary. A discussion of the implications of Theorem 1 and Corollary 1 is presented in Section IV along with some remarks on the practical implementation of an approximately capacity-achieving communication system. Finally, the results and conclusions of the paper are summarized in Section V.

## II. Technical Background

Before stating and proving the main results of the paper, we review the implications of our basic assumption that the communiction range on the channel is much larger than the radius of the sphere containing the transmitting and receiving antennas. In particular, this assumption allows us to convert the original problem, in which the antennas are allowed to be distributed arbitrarily throughout two spherical volumes, to an equivalent problem in which the antennas are constrained to be distributed over the circular surfaces passing through the centers of the original spherical volumes and perpendicular to the line connecting those two centers.

To see that this is true, consider again the situation illustrated in Figure 1, and let $r_T$ and $r_R$ represent the radii of the spherical volumes $\mathcal{V}_T$ and $\mathcal{V}_R$, respectively. In this case, for sufficiently large $d \gg \max(r_T, r_R)$, the distance $d_{ij}$ between any two transmitter-receiver pairs $\{\mathbf{u}_{T,j}, \mathbf{v}_{R,i}\}$ of the form $\mathbf{u}_{T,j} = (x_{T,j}, y_{T,j}, z_{T,j})^T$ and $\mathbf{v}_{R,i} = (x_{R,i}, y_{R,i}, z_{R,i})^T$ is well approximated as

$$\begin{aligned}
d_{ij} &= \left[(x_{R,i} - x_{T,j})^2 + (y_{R,i} - y_{T,j})^2 + (z_{R,i} - z_{T,j})^2\right]^{1/2} \\
&= (x_{R,i} - x_{T,j})\left[1 + \frac{(y_{R,i} - y_{T,j})^2 + (z_{R,i} - z_{T,j})^2}{(x_{R,i} - x_{T,j})^2}\right]^{1/2} \\
&\approx (x_{R,i} - x_{T,j})\left[1 + \frac{(y_{R,i} - y_{T,j})^2 + (z_{R,i} - z_{T,j})^2}{2(x_{R,i} - x_{T,j})^2}\right] \\
&= (x_{R,i} - x_{T,j}) + \frac{(y_{R,i} - y_{T,j})^2 + (z_{R,i} - z_{T,j})^2}{2(x_{R,i} - x_{T,j})}.
\end{aligned} \quad (2.1)$$

Hence, the elements of $\mathbf{H}$ take the form

$$\begin{aligned}
h_{ij} &= e^{-i2\pi \frac{d_{ij}}{\lambda}} \approx e^{-i\frac{2\pi}{\lambda}\left[(x_{R,i} - x_{T,j}) + \frac{(y_{R,i} - y_{T,j})^2 + (z_{R,i} - z_{T,j})^2}{2d}\right]} \\
&= e^{-i\frac{2\pi}{\lambda}(x_{R,i} - x_{T,j})} e^{-i\frac{\pi}{\lambda d}\left[(y_{R,i}^2 + z_{R,i}^2) + (y_{T,j}^2 + z_{T,j}^2)\right]} e^{i2\pi \frac{(y_{R,i} y_{T,j}) + (z_{R,i} z_{T,j})}{\lambda d}},
\end{aligned} \quad (2.2)$$

and $\mathbf{H}$ can be rewritten as

$$\mathbf{H} \approx (\mathbf{h}_R \mathbf{h}_T^*) \circ \tilde{\mathbf{H}}, \quad (2.3)$$

where the notation $\mathbf{A} \circ \mathbf{B}$ denotes the Hadamard (i.e., element-wise) product of the matrices $\mathbf{A}$ and $\mathbf{B}$ and

$$\mathbf{h}_T = \left( e^{i\frac{2\pi}{\lambda}\left(x_{T,1} - \frac{1}{2d}y_{T,1}^2 - \frac{1}{2d}z_{T,1}^2\right)}, e^{i\frac{2\pi}{\lambda}\left(x_{T,2} - \frac{1}{2d}y_{T,2}^2 - \frac{1}{2d}z_{T,2}^2\right)}, \ldots, e^{i\frac{2\pi}{\lambda}\left(x_{T,M} - \frac{1}{2d}y_{T,M}^2 - \frac{1}{2d}z_{T,M}^2\right)} \right)^*,$$

$$\mathbf{h}_R = \left( e^{i\frac{2\pi}{\lambda}\left(x_{R,1} + \frac{1}{2d}y_{R,1}^2 + \frac{1}{2d}z_{R,1}^2\right)}, e^{i\frac{2\pi}{\lambda}\left(x_{R,2} + \frac{1}{2d}y_{R,2}^2 + \frac{1}{2d}z_{R,2}^2\right)}, \ldots, e^{i\frac{2\pi}{\lambda}\left(x_{R,M} + \frac{1}{2d}y_{R,M}^2 + \frac{1}{2d}z_{R,M}^2\right)} \right)^*, \quad (2.4)$$

$$\tilde{\mathbf{H}} = \left( \tilde{h}_{ij} \right) = \left( e^{i 2\pi \frac{\left(y_{R,i} y_{T,j}\right) + \left(z_{R,i} z_{T,j}\right)}{\lambda d}} \right).$$

Note that the matrix $\tilde{\mathbf{H}}$ depends only on the coordinates of the transmitter-receiver pairs $\{\mathbf{u}_{T,j}, \mathbf{v}_{R,i}\}$ projected onto the circular regions in the $y$-$z$ plane of the local coordinate system denoted by $\mathcal{S}_T$ and $\mathcal{S}_R$, as represented by the dark blue discs in Figure 1. We have the following useful lemma regarding $\mathbf{H}$ and $\tilde{\mathbf{H}}$.

**Lemma 1.** The singular values of $\mathbf{H}$ are equivalent to the singular values of $\tilde{\mathbf{H}}$.

**Proof.** Let $\tilde{\mathbf{H}} = \mathbf{U} \mathbf{\Lambda} \mathbf{V}^*$ represent the singular value decomposition (SVD) of the matrix $\tilde{\mathbf{H}}$. It follows from Equation (2.3) that

$$\mathbf{H} = \text{diag}(\mathbf{h}_R) \tilde{\mathbf{H}} \text{diag}(\bar{\mathbf{h}}_T) = \text{diag}(\mathbf{h}_R) \mathbf{U} \mathbf{\Lambda} \mathbf{V}^* \text{diag}(\bar{\mathbf{h}}_T) = \mathbf{W} \mathbf{\Lambda} \mathbf{Z}^*, \quad (2.5)$$

where an overbar indicates complex conjugation

$$\begin{aligned} \mathbf{W} &= \text{diag}(\mathbf{h}_R) \mathbf{U}, \\ \mathbf{Z} &= \text{diag}(\mathbf{h}_T) \mathbf{V}, \end{aligned} \quad (2.6)$$

and

$$\begin{aligned} \mathbf{W}^* \mathbf{W} &= \mathbf{U}^* \text{diag}(\bar{\mathbf{h}}_R) \text{diag}(\mathbf{h}_R) \mathbf{U} = \mathbf{U}^* \mathbf{U} = \mathbf{I}, \\ \mathbf{Z}^* \mathbf{Z} &= \mathbf{V}^* \text{diag}(\bar{\mathbf{h}}_T) \text{diag}(\mathbf{h}_T) \mathbf{V} = \mathbf{V}^* \mathbf{V} = \mathbf{I}. \end{aligned} \quad (2.7)$$

Hence, $\mathbf{H} = \mathbf{W} \mathbf{\Lambda} \mathbf{Z}^*$ represents a version of the SVD of $\mathbf{H}$, and the singular values of $\mathbf{H}$ are equivalent to the singular values of $\tilde{\mathbf{H}}$. ∎

It follows that to determine the maximum achievable spectral efficiency $\xi_M(\mathbf{H})$ for the problem illustrated in Figure 1, we can really solve a completely equivalent problem that depends entirely on antenna distribution in only two dimensions ($y$ and $z$) rather than three, in which the matrix $\mathbf{H}$ is replaced by $\tilde{\mathbf{H}}$. For this equivalent problem, the capacity expressions given by Equations (1.4) and (1.6) remain exactly the same as the original problem since the singular values of $\mathbf{H}$ are equivalent to the singular values of $\tilde{\mathbf{H}}$.

As we discuss below, this equivalence implies that the coupling associated with the distributed transmitter and receiver antennas in the more general problem of interest in this paper is very accurately represented by a two-dimensional spatial Fourier transform, in which the spatial signal is associated with the transmitter, and the corresponding spatial-frequency signal (conceptually scaled by $d$ and measured in radians of azimuth and elevation) is associated with the receiver. The equivalent Fourier-domain problem is illustrated in Figure 2. Note that the three-dimensional transmitter and receiver coordinates $\mathbf{u}_{T,j}$ and $\mathbf{v}_{R,i}$ have been replaced by the two-dimensional equivalents $\tilde{\mathbf{u}}_{T,j} = (y_{T,j}, z_{T,j})^T$ and $\tilde{\mathbf{v}}_{R,i} = (y_{R,i}, z_{R,i})^T$.

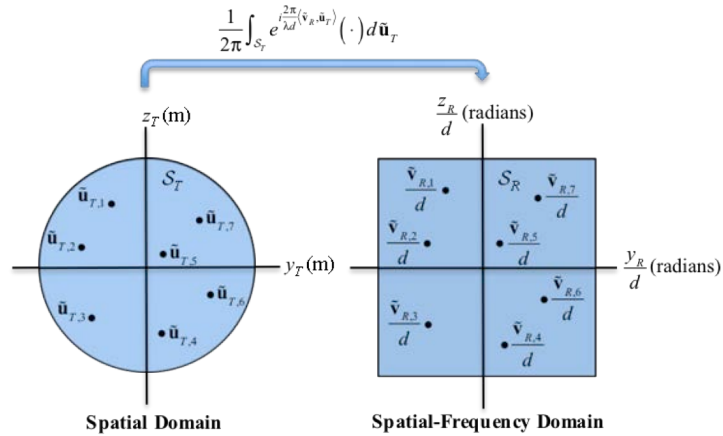

Figure 2. Equivalent 2-D spatial Fourier transform representation.

### III. MAIN RESULTS

Throughout the remainder of the paper, we assume (without loss of generality) that $\mathcal{S}_T$ and $\mathcal{S}_R$ are each discs with identical radius $R$ and area $|\mathcal{S}| = \pi R^2$, centered at the information source (transmitter) and the information sink (receiver), respectively. We abuse notation slighly by writing $\mathcal{S} = \mathcal{S}_T = \mathcal{S}_R$, where the necessary translation to the appropriate center is understood. We also assume that $d \gg R$ (equivalently, $|\mathcal{S}| \ll \pi d^2$) and that $\sqrt{(A_T A_R)/(\lambda^2 d^2)} \ll 1$. Recalling the discussion in the previous section, let us consider the channel model given by (with the time variable suppressed, as usual)

$$y(\mathbf{v}) = \int_{\mathcal{A}_T} H(\mathbf{v},\mathbf{u}) x(\mathbf{u}) d\mathbf{u} + n(\mathbf{v}), \quad \mathbf{v} \in \mathcal{A}_R, \tag{3.1}$$

where $\mathcal{A}_T$ and $\mathcal{A}_R$ are arbitrary *admissible aperture sets* that satisfy the aperture constraints

$$\begin{aligned} |\mathcal{A}_T| &= A_T, \\ |\mathcal{A}_R| &= A_R, \end{aligned} \tag{3.2}$$

and

$$H(\mathbf{v},\mathbf{u}) = \sqrt{\frac{L}{\lambda^2 d^2}} e^{i\frac{2\pi}{\lambda d}\langle \mathbf{v},\mathbf{u}\rangle}, \tag{3.3}$$

for $\mathbf{u}, \mathbf{v} \in \mathcal{S}$. Hence, the transmitted symbols may be radiated with varying intensity and phase from an arbitrary set of points $\mathcal{A}_T \subset \mathcal{S}_T$ with aperture area $A_T$, and the received signal is observed over an arbitrary set of points $\mathcal{A}_R \subset \mathcal{S}_R$ with aperture area $A_R$. Let $n(\mathbf{v})$ be a complex Gaussian white noise process on $\mathcal{S}$ with power spectral density $N_0$ with respect to (wrt) the suppressed time variable, and let the input baseband information process $x(\mathbf{u})$ be bandlimited to bandwidth $B$ wrt temporal frequency and satisfy the power constraint

$$E\left\{\int_{\mathcal{A}_T} |x(\mathbf{u})|^2 d\mathbf{u}\right\} \leq P, \tag{3.4}$$

To make use of this model, we note that the kernel $H(\mathbf{v},\mathbf{u}): \mathcal{S} \to \mathcal{S}$ defines a compact normal operator $\mathcal{H}_\mathcal{S}$ mapping the Hilbert space $\mathcal{L}^2(\mathcal{S})$ of square-integrable functions on $\mathcal{S}$ into itself. As such it can be represented as

$$H(\mathbf{v},\mathbf{u}) = \sum_{n=1}^{\infty} v_n p_n(\mathbf{v}) \bar{p}_n(\mathbf{u}), \tag{3.5}$$

where $\{v_n\}_{n=1}^{\infty}$ represents the sequence of eigenvalues for the operator $\mathcal{H}_\mathcal{S}$ satisfying

$$|v_1| \geq |v_2| \geq |v_3| \geq \ldots, \tag{3.6}$$

and

$$\|\mathcal{H}_\mathcal{S}\|^2 = \sum_{n=1}^{\infty} |v_n|^2 = \int_\mathcal{S} \int_\mathcal{S} |H(\mathbf{v},\mathbf{u})|^2 d\mathbf{u} d\mathbf{v} = \frac{L}{\lambda^2 d^2} |\mathcal{S}|^2, \tag{3.7}$$

and $\{p_n\}_{n=1}^{\infty}$ is a sequence of orthonormal functions that represent the eigenfunctions for $\mathcal{H}_S$. Furthermore, either finitely many of the $\{v_n\}_{n=1}^{\infty}$ are nonzero or $|v_n| \to 0$ as $n \to \infty$, and the set $\{p_n\}_{n=1}^{\infty}$ spans $\mathcal{L}^2(\mathcal{S})$ if and only if $|v_n| > 0$, for all $n = 1, 2, \ldots$, [22].

Similarly, the restriction of the operator $\mathcal{H}_S$ to arbitrary admissible aperture sets $\mathcal{A}_T$ and $\mathcal{A}_R$, which we denote by $\mathcal{H}_{\mathcal{A}_T:\mathcal{A}_R}$, is a compact operator defined by the kernel $H_{\mathcal{A}_T:\mathcal{A}_R}(\mathbf{v},\mathbf{u}): \mathcal{A}_T \to \mathcal{A}_R$ given by

$$H_{\mathcal{A}_T:\mathcal{A}_R}(\mathbf{v},\mathbf{u}) = \begin{cases} H(\mathbf{v},\mathbf{u}), & \mathbf{u} \in \mathcal{A}_T, \mathbf{v} \in \mathcal{A}_R, \\ 0, & \text{otherwise.} \end{cases}$$

that maps $\mathcal{L}^2(\mathcal{A}_T)$ into $\mathcal{L}^2(\mathcal{A}_R)$. Formally, we have $\mathcal{H}_{\mathcal{A}_T:\mathcal{A}_R} = \mathcal{P}_{\mathcal{A}_R} \mathcal{H} \mathcal{P}_{\mathcal{A}_T}$, where $\mathcal{P}_{\mathcal{A}_T}$ and $\mathcal{P}_{\mathcal{A}_R}$ are the projection operators onto the spaces $\mathcal{L}^2(\mathcal{A}_T)$ and $\mathcal{L}^2(\mathcal{A}_R)$, respectively. The properties of $\mathcal{H}_{\mathcal{A}_T:\mathcal{A}_R}$ are very similar to those for $\mathcal{H}_S$. That is [23],

$$H_{\mathcal{A}_T:\mathcal{A}_R}(\mathbf{v},\mathbf{u}) = \sum_{n=1}^{\infty} \eta_n r_n(\mathbf{v}) \bar{q}_n(\mathbf{u}), \qquad (3.8)$$

where $\{\eta_n\}_{n=1}^{\infty}$ represents the sequence of singular values for the operator $\mathcal{H}_{\mathcal{A}_T:\mathcal{A}_R}$ satisfying

$$|\eta_1| \geq |\eta_2| \geq |\eta_3| \geq \ldots, \qquad (3.9)$$

and

$$\left\| \mathcal{H}_{\mathcal{A}_T:\mathcal{A}_R} \right\|^2 = \sum_{n=1}^{\infty} |\eta_n|^2 = \int_{\mathcal{S}} \int_{\mathcal{S}} \left| H_{\mathcal{A}_T:\mathcal{A}_R}(\mathbf{u},\mathbf{v}) \right|^2 d\mathbf{u} d\mathbf{v} = \frac{A_T A_R L}{\lambda^2 d^2} \ll 1. \qquad (3.10)$$

In this case, $\{q_n\}_{n=1}^{\infty}$ and $\{r_n\}_{n=1}^{\infty}$ are sequences of orthonormal functions that represent the singular functions for $\mathcal{H}_{\mathcal{A}_T:\mathcal{A}_R}$, and the sets $\{q_n\}_{n=1}^{\infty}$ and $\{r_n\}_{n=1}^{\infty}$ span $\mathcal{L}^2(\mathcal{A}_T)$ and $\mathcal{L}^2(\mathcal{A}_R)$, respectively if and only if $|\eta_n| > 0$, for all $n = 1, 2, \ldots$.

For the operator $\mathcal{H}_S$ considered here, which is really just a two-dimensional spatial Fourier transform operator, the eigenfunctions $\{p_n\}_{n=1}^{\infty}$ are versions of the two-dimensional prolate spheroidal wave functions, which, along with the eigenvalues $\{v_n\}_{n=1}^{\infty}$, satisfy the following properties [24, 25]:

P1. $|v_n| > 0, \quad \forall n = 1, 2, \ldots,$

P2. $|v_1|^2 \approx |v_2|^2 \approx \cdots \approx |v_\mathcal{M}|^2 \approx L < 1$ and $\sum_{n=\mathcal{M}+1}^{\infty} |v_n|^2 \approx 0$, where $\mathcal{M} = \left\lceil |\mathcal{S}|^2 / \lambda^2 d^2 \right\rceil$ for $|\mathcal{S}|^2 / \lambda^2 d^2 \geq 1$,

P3. $\{p_n\}_{n=1}^{\mathcal{M}}$ are essentially space- and band-limited to the set $\mathcal{S}$ in both domains; that is, $p_n(\mathbf{u}) = 0$ for $\mathbf{u} \notin \mathcal{S}$, and $(\mathcal{H}_\mathcal{S} p_n)(\mathbf{v}) \approx 0$ for $\mathbf{v} \notin \mathcal{S}$.

For the sake of completeness, the exact forms of $\{v_n\}_{n=1}^{\infty}$ and $\{p_n\}_{n=1}^{\infty}$ are given in Appendix A.

It should be noted that for the restricted operator $\mathcal{H}_{A_T : A_R}$, for which we have $\left\| \mathcal{H}_{A_T : A_R} \right\|^2 = (A_T A_R L) / (\lambda^2 d^2) \ll 1$, the singular values $\{\eta_n\}_{n=1}^{\infty}$, will not necessarily possess the "nice" features of Property P2 above. In particular, it will not necessarily be the case that $|\eta_1|^2 \approx |\eta_2|^2 \approx \cdots \approx |\eta_\mathcal{M}|^2 \approx c$ for some constant $c > 0$ and some integer $\mathcal{M} \geq 1$, with $\sum_{n=\mathcal{M}+1}^{\infty} |\eta_n|^2 \approx 0$. In fact, if the sets $A_T$ and $A_R$ are chosen carefully, that will indeed be the case.

Now, expanding the transmitted information process wrt the basis $\{q_n\}_{n=1}^{\infty}$ for an abritrary admissible aperture set $A_T$ gives

$$x(\mathbf{u}) = \sum_{n=1}^{\infty} x_n q_n(\mathbf{u}), \qquad (3.11)$$

subject to the power constraint

$$E\left\{ \sum_{n=1}^{\infty} |x_n|^2 \right\} \leq P. \qquad (3.12)$$

Similarly, the output process may be expanded in terms of the basis $\{r_n\}_{n=1}^{\infty}$ for an arbitrary admissible aperture set $A_R$ as

$$y(\mathbf{v}) = \sum_{n=1}^{\infty} y_n r_n(\mathbf{v}), \quad (3.13)$$

and the channel model given by Equation (3.1) becomes

$$y(\mathbf{v}) = \int_{A_T} H(\mathbf{v},\mathbf{u}) x(\mathbf{u}) d\mathbf{u} + n(\mathbf{v}) = \sum_{n=1}^{\infty} (\eta_n x_n + n_n) r_n(\mathbf{v}) = \sum_{n=1}^{\infty} y_n r_n(\mathbf{v}) \quad \mathbf{v} \in \mathcal{A}_R. \quad (3.14)$$

Equivalently, in vector notation representing the expansion wrt to the dual basis $\{q_n\}_{n=1}^{\infty}$, $\{r_n\}_{n=1}^{\infty}$ at an arbitrary observation time, the channel model becomes

$$\mathbf{y} = \mathbf{\eta} \circ \mathbf{x} + \mathbf{n}, \quad (3.15)$$

where $\mathbf{y} = \{y_n\}_{n=1}^{\infty}$, $\mathbf{\eta} = \{\eta_n\}_{n=1}^{\infty}$, $\mathbf{x} = \{x_n\}_{n=1}^{\infty}$, and $\mathbf{n} = \{n_k\}_{k=1}^{\infty}$. With respect to the suppressed time variable, the $\{n_k\}_{k=1}^{\infty}$ are now iid complex white Gaussian noise processes with power spectral density $N_0$. Furthermore, given any set of individual power levels $\{P_n\}_{n=1}^{\infty}$ that satisfy

$$\left\{ P_n \geq 0 \middle| \sum_{n=1}^{\infty} P_n \leq P, P_n \equiv 0 \ \forall n > N, \text{ for some integer } N > 0 \right\}_{n=1}^{\infty}, \quad (3.16)$$

the maximum achievable spectral efficiency for Channel (3.15) subject to the set of constraints

$$E\{|x_n|^2\} = P_n, \quad n = 1, 2, \ldots, \quad (3.17)$$

is given by [19, (9.62) and (9.71)]

$$\xi_{\{\eta_n, P_n\}} = \sum_{n=1}^{\infty} \log_2 \left( 1 + |\eta_n|^2 \frac{P_n}{BN_0} \right). \quad (3.18)$$

We can now state and prove the main result of this paper.

**Theorem 1.** Let $A_T$, $A_R$, $\lambda$, $d$, $B$, $N_0$, $g$, $\gamma$, and $\mathcal{S} = \mathcal{S}_T = \mathcal{S}_R$ be as discussed previously. Define the maximum achievable spectral efficiency $\xi_{\gamma g}$ for the class of all channels of the form (3.1) subject to aperture constraints (3.2) and total power constraint (3.3) as[1]

---

[1] We conjecture that $\xi_{\gamma g}$ as defined here is the true information capacity of an arbitrary continuous-time, complex-valued, bandlimited, free-space, AWGN channel that satisfies our far field assumption. That is, $\xi_{\gamma g}$ is the information capacity for the class of all channels satisfying Equation (3.1) subject to aperture constraints (3.2) and power constraint (3.3) for arbitrary $\mathcal{S}$. Unfortunately, proof of this conjecture would require proving a corresponding

$$\xi_{\gamma g} = \sup_{\mathcal{S}, \mathcal{A}_T, \mathcal{A}_R, \{P_n\}_{n=1}^{\infty}} \left\{ \xi_{\{\eta_n, P_n\}} \middle| \begin{array}{l} \{\eta_n\}_{n=1}^{\infty} \mathcal{A}_T \subset \mathcal{S}_T, \mathcal{A}_R \subset \mathcal{S}_R, \text{ admissible,} \\ \{P_n\}_{n=1}^{\infty} \text{ satisfying (3.15)} \end{array} \right\}. \quad (3.19)$$

Then $\xi_{\gamma g}$ satisfies

$$\begin{array}{c} \xi_{\gamma g} = \log_2(1+\gamma g), \qquad\qquad\qquad \gamma g \leq \varepsilon_0 - 1, \\ \sum_{k=1}^{K} \log_2\left( \frac{|v_k|^2}{BN_0} \left[ \frac{A_T A_R P}{K|\mathcal{S}|^2} + \frac{1}{K}\sum_{i=1}^{K} \frac{BN_0}{|v_i|^2} \right] \right) \leq \xi_{\gamma g} \leq \sqrt{\frac{\gamma g}{\varepsilon_0 - 1}} \log_2(\varepsilon_0), \quad \gamma g > \varepsilon_0 - 1, \end{array} \quad (3.20)$$

where $\{v_n\}_{n=1}^{\infty}$ are the eigenvalues of the operator $\mathcal{H}_{\mathcal{S}}$ with $K$ and $\mathcal{S}$ chosen such that

$$\max(A_T, A_R) \leq |\mathcal{S}| \ll \pi d^2,$$

$$K = \max\left\{ \kappa \in \mathbb{Z}^+ : \frac{P}{\kappa} + \frac{|\mathcal{S}|^2}{\kappa A_T A_R} \sum_{i=1}^{\kappa} \frac{BN_0}{|v_i|^2} > \frac{|\mathcal{S}|^2 BN_0}{A_T A_R |v_k|^2}, \forall k = 1, 2, \ldots, \kappa \right\} < \infty, \quad (3.21)$$

and $\varepsilon_0 \approx 4.9215$ is the solution of the transcendental equation

$$\varepsilon = e^{2\left(1-\frac{1}{\varepsilon}\right)}, \quad \varepsilon > 1. \quad (3.22)$$

**Proof.** It follows from Equation (3.18) that

$$\xi_{\gamma g} \leq \sup_{\left\{ \sum_{n=1}^{\infty} P_n = P, \sum_{n=1}^{\infty} |\eta_n|^2 = \frac{A_T A_R L}{\lambda^2 d^2} \right\}} \left[ \sum_{n=1}^{\infty} \log_2\left( 1 + |\eta_n|^2 \frac{P_n}{BN_0} \right) \right]. \quad (3.23)$$

Further, given any set $\{\eta_n\}_{n=1}^{\infty}$ such that $\sum_{n=1}^{\infty} |\eta_n|^2 = (A_T A_R L)/(\lambda^2 d^2)$, it is straightforward to show that (see Appendix B)

$$\sup_{\left\{ \sum_{n=1}^{\infty} P_n = P \right\}} \left[ \sum_{n=1}^{\infty} \log_2\left( 1 + |\eta_n|^2 \frac{P_n}{BN_0} \right) \right] = \max_{\left\{ \sum_{n=1}^{\infty} P_n = P \right\}} \left[ \sum_{n=1}^{\infty} \log_2\left( 1 + |\eta_n|^2 \frac{P_n}{BN_0} \right) \right]$$

$$= \sum_{k=1}^{K} \log_2\left( \frac{|\eta_k|^2}{BN_0} \left[ \frac{P}{K} + \frac{1}{K}\sum_{i=1}^{K} \frac{BN_0}{|\eta_i|^2} \right] \right), \quad (3.24)$$

---

coding theorem for an abstract channel representing the class of all such channels, which is beyond the scope of the current paper and is a topic for future research.

where *K* is the greatest integer such that

$$\frac{P}{K} + \frac{1}{K}\sum_{i=1}^{K}\frac{BN_0}{|\eta_i|^2} \geq \frac{BN_0}{|\eta_k|^2}, \quad \forall k = 1, 2, \ldots, K. \tag{3.25}$$

Hence,

$$\sup_{\left\{\sum_{n=1}^{\infty}P_n=P,\sum_{n=1}^{\infty}|\eta_n|^2=\frac{A_T A_R L}{\lambda^2 d^2}\right\}} \left[\sum_{n=1}^{\infty}\log_2\left(1+|\eta_n|^2\frac{P_n}{BN_0}\right)\right]$$

$$= \sup_{\left\{\sum_{n=1}^{\infty}|\eta_n|^2=\frac{A_T A_R L}{\lambda^2 d^2}\right\}} \left[\sum_{k=1}^{K}\log_2\left(\frac{|\eta_k|^2}{BN_0}\left[\frac{P}{K}+\frac{1}{K}\sum_{i=1}^{K}\frac{BN_0}{|\eta_i|^2}\right]\right)\right], \tag{3.26}$$

where *K* is the greatest integer satisfying Equation (3.25). Since $\log_2(\cdot)$ is a strictly convex function, it follows that for any such *K*, we have

$$\sup_{\left\{\sum_{n=1}^{\infty}|\eta_n|^2=\frac{A_T A_R L}{\lambda^2 d^2}\right\}} \left[\sum_{k=1}^{K}\log_2\left(\frac{|\eta_k|^2}{BN_0}\left[\frac{P}{K}+\frac{1}{K}\sum_{i=1}^{K}\frac{BN_0}{|\eta_i|^2}\right]\right)\right]$$

$$\leq \sup_{\left\{\sum_{k=1}^{K}|\eta_k|^2=\frac{A_T A_R L}{\lambda^2 d^2}\right\}} \left[K\log_2\left(\frac{1}{K}\sum_{k=1}^{K}\frac{|\eta_k|^2}{BN_0}\left[\frac{P}{K}+\frac{1}{K}\sum_{i=1}^{K}\frac{BN_0}{|\eta_i|^2}\right]\right)\right]$$

$$= K\log_2\left(1+\frac{A_T A_R LP}{K^2\lambda^2 d^2 BN_0}\right) = K\log_2\left(1+\frac{\gamma g}{K^2}\right) \tag{3.27}$$

$$\leq \begin{cases} \log_2(1+\gamma g), & \gamma g \leq \varepsilon_0 - 1, \\ \max_{x>0}\left\{x\log_2\left(1+\frac{\gamma g}{x^2}\right)\right\}, & \gamma g > \varepsilon_0 - 1, \end{cases} = \begin{cases} \log_2(1+\gamma g), & \gamma g \leq \varepsilon_0 - 1, \\ \sqrt{\frac{\gamma g}{\varepsilon_0 - 1}}\log_2(\varepsilon_0), & \gamma g > \varepsilon_0 - 1. \end{cases}$$

Hence,

$$\xi_{\gamma g} \leq \begin{cases} \log_2(1+\gamma g), & \gamma g \leq \varepsilon_0 - 1, \\ \sqrt{\frac{\gamma g}{\varepsilon_0 - 1}}\log_2(\varepsilon_0), & \gamma g > \varepsilon_0 - 1, \end{cases} \tag{3.28}$$

which establishes the upper bound in Equation (3.20).

To establish the lower bound, recall that we always assume $\sqrt{(A_T A_R)/(\lambda^2 d^2)} \ll 1$. Hence, if we choose

$$\begin{aligned} \mathcal{A}_T &= \left\{ \mathbf{u} = (x_T, y_T) \Big| x_T^2 + y_T^2 \leq \frac{A_T}{\pi} \right\}, \\ \mathcal{A}_R &= \left\{ \mathbf{v} = (x_R, y_R) \Big| x_R^2 + y_R^2 \leq \frac{A_R}{\pi} \right\}, \end{aligned} \quad (3.29)$$

then

$$H_{\mathcal{A}_T : \mathcal{A}_R}(\mathbf{v}, \mathbf{u}) \approx \begin{cases} \sqrt{\dfrac{L}{\lambda^2 d^2}}, & \mathbf{u} \in \mathcal{A}_T, \mathbf{v} \in \mathcal{A}_R, \\ 0, & \text{otherwise.} \end{cases} \quad (3.30)$$

Clearly, if we let $\{q_n\}_{n=1}^{\infty}$ be any orthonormal basis for $\mathcal{A}_T$ for which $q_1(\mathbf{u}) \equiv 1/\sqrt{A_T}$ and $\{r_n\}_{n=1}^{\infty}$ be any orthonormal basis for $\mathcal{A}_R$ for which $r_1(\mathbf{v}) \equiv 1/\sqrt{A_R}$, then $\mathcal{H}_{\mathcal{A}_T : \mathcal{A}_R}$ can be represented as in Equation (3.8), where $\eta_1 = \sqrt{(A_T A_R L)/(\lambda^2 d^2)}$ and $\eta_n = 0$ for all $n = 2, 3, \ldots$. Hence, for this choice of admissible aperture sets, Equation (3.18) becomes

$$\xi_{\{\eta_n, P_n\}} = \log_2\left(1 + \frac{A_T A_R L P}{\lambda^2 d^2 B N_0}\right) = \log_2(1 + \gamma g), \quad (3.31)$$

and it follows that $\xi_{\gamma g} \geq \log_2(1 + \gamma g)$ for all values of $\gamma g > 0$. This establishes Equation (3.20) for the case $\gamma g \leq \varepsilon_0 - 1$.

To establish the lower bound for $\gamma g > \varepsilon_0 - 1$, it is sufficient to show that for any values of $|\mathcal{S}|$ and $K$ satisfying Equation (3.21), there exists a sequence of sets of functions $\{Q_N = \{q_{N,k}(\mathbf{u})\}_{k=1}^{K}\}_{N=1}^{\infty}$ defined on a sequence of admissible aperture sets $\{\mathcal{A}_{T,N}\}_{N=1}^{\infty}$ and a corresponding sequence of sets of functions $\{R_N = \{r_{N,k}(\mathbf{v})\}_{k=1}^{K}\}_{N=1}^{\infty}$ defined on a sequence of admissible aperture sets $\{\mathcal{A}_{R,N}\}_{N=1}^{\infty}$ such that the sequence of maximum achievable spectral efficiencies $\xi_{\{Q_N, R_N, P_N\}}$ for the sequence of observational models

$$y_{N,1} = x_{N,1} \int_{\mathcal{A}_{T,N}} \bar{r}_{N,1}(\mathbf{v}) \int_{\mathcal{A}_{R,n}} H(\mathbf{v},\mathbf{u}) q_{N,1}(\mathbf{u}) d\mathbf{u} d\mathbf{v} + \int_{\mathcal{A}_{R,N}} \bar{r}_{N,1}(\mathbf{v}) n(\mathbf{v}) d\mathbf{v},$$

$$y_{N,2} = x_{N,2} \int_{\mathcal{A}_{T,N}} \bar{r}_{N,2}(\mathbf{v}) \int_{\mathcal{A}_{R,N}} H(\mathbf{v},\mathbf{u}) q_{N,2}(\mathbf{u}) d\mathbf{u} d\mathbf{v} + \int_{\mathcal{A}_{R,n}} \bar{r}_{N,2}(\mathbf{v}) n(\mathbf{v}) d\mathbf{v},$$

$$\vdots$$

$$y_{N,K} = x_{N,K} \int_{\mathcal{A}_{T,N}} \bar{r}_{N,K}(\mathbf{v}) \int_{\mathcal{A}_{R,N}} H(\mathbf{v},\mathbf{u}) q_{N,K}(\mathbf{u}) d\mathbf{u} d\mathbf{v} + \int_{\mathcal{A}_{R,N}} \bar{r}_{N,K}(\mathbf{v}) n(\mathbf{v}) d\mathbf{v},$$

(3.32)

subject to the sequence of constraints $\left\{ P_N = \left\{ P_{N,k} \right\}_{k=1}^{K} \right\}_{N=1}^{\infty}$ satisfying

$$E\left\{ \left| x_{N,k} \right|^2 \right\} = P_{N,k}, \quad \sum_{k=1}^{K} P_{N,k} \leq P, \quad N = 1, 2, \ldots,$$

(3.33)

converges to the lower bound in Equation (3.20). That is,

$$\xi_{\{Q_N, R_N, P_N\}} \xrightarrow[N \to \infty]{} \sum_{k=1}^{K} \log_2 \left( \frac{|v_k|^2}{BN_0} \left[ \frac{A_T A_R P}{K |\mathcal{S}|^2} + \frac{1}{K} \sum_{i=1}^{K} \frac{BN_0}{|v_i|^2} \right] \right).$$

(3.34)

This conditions is sufficient since for any $\varepsilon > 0$, we can then find $N > 0$ and a collection $\xi_{\{Q_N, R_N, P_N\}}$ such that

$$\xi_{\{Q_N, R_N, P_N\}} > \sum_{k=1}^{K} \log_2 \left( \frac{|v_k|^2}{BN_0} \left[ \frac{A_T A_R P}{K |\mathcal{S}|^2} + \frac{1}{K} \sum_{i=1}^{K} \frac{BN_0}{|v_i|^2} \right] \right) - \varepsilon.$$

(3.35)

Hence, we must have

$$\xi_{\gamma g} \geq \sup_{N > 0} \xi_{\{Q_N, R_N, P_N\}} \geq \sum_{k=1}^{K} \log_2 \left( \frac{|v_k|^2}{BN_0} \left[ \frac{A_T A_R P}{K |\mathcal{S}|^2} + \frac{1}{K} \sum_{i=1}^{K} \frac{BN_0}{|v_i|^2} \right] \right),$$

(3.36)

as claimed.

Towards this end, let $|\mathcal{S}|$ and $K$ satisfying Equation (3.21) be given, and let $\{p_i\}_{i=1}^{\infty}$ be the set of eigenfunctions for the operator $\mathcal{H}_\mathcal{S}$ defined on $\mathcal{S}$ with corresponding eigenvalues $\{v_i\}_{i=1}^{\infty}$. Since the functions $\{p_i\}_{i=1}^{\infty}$ are eigenfunctions of the kernel $H(\mathbf{v},\mathbf{u})$, which is continuous on the set $\mathcal{S} \times \mathcal{S}$, they themselves are continuous on the set $\mathcal{S}$. Furthermore, since $\mathcal{S}$ is compact and $K$ is finite, all of the functions $H(\mathbf{v},\mathbf{u})$, $\{p_i\}_{i=1}^{K}$, and

$$\left\{ \overline{p}_i(\mathbf{v}) H(\mathbf{v},\mathbf{u}) p_j(\mathbf{u}) \,\middle|\, \mathbf{v},\mathbf{u} \in \mathcal{S}, i,j = 1,2,\ldots,K \right\},$$

are continuous, and the continuity is uniform for all $\mathbf{v},\mathbf{u} \in \mathcal{S}$ and all $i,j = 1,2,\ldots,K$. Now let $\left\{ \mathcal{S}_N^T \right\}_{N=1}^{\infty}$ be a sequence of partitions of the set $\mathcal{S}_T = \mathcal{S}$ such that

$$\mathcal{S}_N^T = \left\{ \mathcal{S}_{N,i}^T \subset \mathcal{S}_T \,\middle|\, \mathcal{S}_T = \bigcup_{i=1}^{K_N^T} \mathcal{S}_{N,i}^T,\ \mathcal{S}_{N,i}^T \cap \mathcal{S}_{N,j}^T = \varnothing, i \neq j,\ \left| \mathcal{S}_{N,i}^T \right| = \frac{|\mathcal{S}|}{N} \right\},\ N = 1,2,\ldots,\infty, \quad (3.37)$$

Also let $\mathcal{B}_\delta(\mathbf{u})$ be a disk of area $\delta$ centered at the origin in $\mathcal{S}$, and let $\left\{ \mathcal{U}_N \right\}_{N=1}^{\infty}$ be a corresponding sequence of sets such that

$$\mathcal{U}_N = \left\{ \mathbf{u}_{N,i} \in \mathcal{S}_T \,\middle|\, \mathcal{B}_{A_T/N}(\mathbf{u} - \mathbf{u}_{N,i}) \subset \mathcal{S}_{N,i}^T,\ i = 1,2,\ldots,K_N^T \right\}. \quad (3.38)$$

Similarly, let $\left\{ \mathcal{S}_N^R \right\}_{N=1}^{\infty}$ be a sequence of partitions of the set $\mathcal{S}_R = \mathcal{S}$ such that

$$\mathcal{S}_N^R = \left\{ \mathcal{S}_{N,i}^R \subset \mathcal{S}_R \,\middle|\, \mathcal{S} = \bigcup_{i=1}^{K_N^R} \mathcal{S}_{N,i}^R,\ \mathcal{S}_{N,i}^R \cap \mathcal{S}_{N,j}^R = \varnothing, i \neq j,\ \left| \mathcal{S}_{N,i}^R \right| = \frac{|\mathcal{S}|}{N} \right\},\ N = 1,2,\ldots,\infty, \quad (3.39)$$

and let $\left\{ \mathcal{V}_N \right\}_{N=1}^{\infty}$ be a corresponding sequence of sets such that

$$\mathcal{V}_N = \left\{ \mathbf{v}_{N,i} \in \mathcal{S}_R \,\middle|\, \mathcal{B}_{A_R/N}(\mathbf{v} - \mathbf{v}_{N,i}) \subset \mathcal{S}_{N,i}^R,\ i = 1,2,\ldots,K_N^R \right\}. \quad (3.40)$$

Then,

$$\begin{aligned}
&\int_\mathcal{S} \overline{\hat{p}_{N,m}^R}(\mathbf{v}) \int_\mathcal{S} H(\mathbf{v},\mathbf{u}) \hat{p}_{N,n}^T(\mathbf{u}) \, d\mathbf{u}\, d\mathbf{v} \\
&= \int_\mathcal{S} \left( \sum_{i=1}^{K_N^R} \overline{p_m(\mathbf{v}_{N,i})} I_{\mathcal{S}_{N,i}^R}(\mathbf{v}) \right) \int_\mathcal{S} H(\mathbf{v},\mathbf{u}) \left( \sum_{j=1}^{K_N^T} p_n(\mathbf{u}_{N,j}) I_{\mathcal{S}_{N,j}^T}(\mathbf{u}) \right) d\mathbf{u}\, d\mathbf{v} \quad (3.41) \\
&\xrightarrow[N \to \infty]{} \int_\mathcal{S} \overline{p_m(\mathbf{v})} \int_\mathcal{S} H(\mathbf{v},\mathbf{u}) p_n(\mathbf{u}) \, d\mathbf{u}\, d\mathbf{v} = v_n \delta_{mn},
\end{aligned}$$

uniformly for all $n,m = 1,2,\ldots,K$, where $\delta_{mn}$ represents the Kronecker delta function, $I_\mathcal{A}(\mathbf{u})$ represents the indicator function for the set $\mathcal{A}$, and

$$\begin{aligned}
\hat{p}_{N,n}^T(\mathbf{u}) &= \sum_{i=1}^{K_N^T} p_n(\mathbf{u}_{N,i}) I_{\mathcal{S}_{N,i}^T}(\mathbf{u}),\quad n = 1,2,\ldots,K, \\
\hat{p}_{N,m}^R(\mathbf{v}) &= \sum_{i=1}^{K_N^R} p_m(\mathbf{v}_{N,i}) I_{\mathcal{S}_{N}^R}(\mathbf{v}),\quad m = 1,2,\ldots,K,
\end{aligned} \quad (3.42)$$

are sets of different simple-function approximations of $\{p_k\}_{k=1}^{K}$, one set defined on the transmitter aperture set and one defined on the receiver aperture set.

If we now let

$$f_{N,n}^{T}(\mathbf{u}) = \sqrt{\frac{|\mathcal{S}|}{A^T}} \sum_{i=1}^{K_N^T} p_n(\mathbf{u}_{N,i}) \mathcal{B}_{A_T/N}(\mathbf{u} - \mathbf{u}_{N,i}), \quad \mathcal{A}_{T,N} = \{\mathbf{u} \in S_T | f_{N,n}^{T}(\mathbf{u}) \neq 0\},$$

$$g_{N,m}^{R}(\mathbf{v}) = \sqrt{\frac{|\mathcal{S}|}{A^R}} \sum_{j=1}^{K_N^R} p_m(\mathbf{v}_{N,j}) \mathcal{B}_{A_R/N}(\mathbf{v} - \mathbf{v}_{N,j}), \quad \mathcal{A}_{R,N} = \{\mathbf{v} \in S_R | g_{N,m}^{R}(\mathbf{v}) \neq 0\},$$

(3.43)

for $n, m = 1, 2, \ldots, K$, then $\mathcal{A}_{T,N}$ and $\mathcal{A}_{R,N}$ are admissible aperture sets, and $\{f_{N,n}^{T}(\mathbf{u})\}_{n=1}^{K}$ and $\{g_{N,m}^{R}(\mathbf{v})\}_{m=1}^{K}$ are unit normal functions defined on $\mathcal{A}_{T,N}$ and $\mathcal{A}_{R,N}$, respectively, derived directly from $\{\hat{p}_{N,n}^{T}(\mathbf{u})\}_{n=1}^{K}$ and $\{\hat{p}_{N,m}^{R}(\mathbf{u})\}_{n=1}^{K}$. We refer to the sets of functions $\{f_{N,n}^{T}(\mathbf{u})\}_{n=1}^{K}$ and $\{g_{N,m}^{R}(\mathbf{v})\}_{m=1}^{K}$ as *admissible distributed antenna functions* to reflect the facts that they are supported on admissible aperture sets and are unit norm. It follows that

$$\lim_{N \to \infty} \int_{\mathcal{S}} \overline{g_{N,m}^{R}(\mathbf{v})} \int_{\mathcal{S}} H(\mathbf{v}, \mathbf{u}) f_{N,n}^{T}(\mathbf{u}) d\mathbf{u} d\mathbf{v}$$

$$= \lim_{N \to \infty} \int_{\mathcal{S}} \left( \sqrt{\frac{|\mathcal{S}|}{A_R}} \sum_{i=1}^{K_N^R} \overline{p_m(\mathbf{v}_{N,i})} \mathcal{B}_{A_R/N}(\mathbf{v} - \mathbf{v}_{N,i}) \right) \int_{\mathcal{S}} H(\mathbf{v}, \mathbf{u}) \left( \sqrt{\frac{|\mathcal{S}|}{A_T}} \sum_{j=1}^{K_N^T} p_n(\mathbf{u}_{N,j}) \mathcal{B}_{A_T/N}(\mathbf{u} - \mathbf{u}_{N,j}) \right) d\mathbf{u} d\mathbf{v}$$

$$= \lim_{N \to \infty} \sqrt{\frac{|\mathcal{S}|^2}{A_T A_R}} \sum_{i=1}^{K_N^R} \sum_{j=1}^{K_N^T} \overline{p_m(\mathbf{v}_{N,i})} p_n(\mathbf{u}_{N,j}) \int_{\mathcal{S}} \int_{\mathcal{S}} \left( \mathcal{B}_{A_R/N}(\mathbf{v} - \mathbf{v}_{N,i}) \right) H(\mathbf{v}, \mathbf{u}) \mathcal{B}_{A_T/N}(\mathbf{u} - \mathbf{u}_{N,j}) d\mathbf{u} d\mathbf{v}$$

$$= \lim_{N \to \infty} \sqrt{\frac{|\mathcal{S}|^2}{A_T A_R}} \sum_{i=1}^{K_N^R} \sum_{j=1}^{K_N^T} \overline{p_m(\mathbf{v}_{N,i})} p_n(\mathbf{u}_{N,j}) \int_{\mathcal{S}} \int_{\mathcal{S}} \left( \mathcal{B}_{A_R/N}(\mathbf{v} - \mathbf{v}_{N,i}) \right) H(\mathbf{v}_{N,i}, \mathbf{u}_{N,j}) \mathcal{B}_{A_T/N}(\mathbf{u} - \mathbf{u}_{N,j}) d\mathbf{u} d\mathbf{v}$$

$$= \lim_{N \to \infty} \sqrt{\frac{A_T A_R}{|\mathcal{S}|^2}} \sum_{i=1}^{K_N^R} \sum_{j=1}^{K_N^T} \overline{p_m(\mathbf{v}_{N,i})} p_n(\mathbf{u}_{N,j}) \int_{\mathcal{S}} \int_{\mathcal{S}} I_{N,i}^{R}(\mathbf{v}) H(\mathbf{v}_{N,i}, \mathbf{u}_{N,j}) I_{N,j}^{T}(\mathbf{u}) d\mathbf{u} d\mathbf{v}$$

(3.44)

$$= \lim_{N \to \infty} \sqrt{\frac{A_T A_R}{|\mathcal{S}|^2}} \sum_{i=1}^{K_N^R} \sum_{j=1}^{K_N^T} \overline{p_m(\mathbf{v}_{N,i})} p_n(\mathbf{u}_{N,j}) \int_{\mathcal{S}} \int_{\mathcal{S}} I_{N,i}^{R}(\mathbf{v}) H(\mathbf{v}, \mathbf{u}) I_{N,j}^{T}(\mathbf{u}) d\mathbf{u} d\mathbf{v}$$

$$= \lim_{N \to \infty} \sqrt{\frac{A_T A_R}{|\mathcal{S}|^2}} \int_{\mathcal{S}} \overline{\hat{p}_{N,m}^{R}(\mathbf{v})} \int_{\mathcal{S}} H(\mathbf{v}, \mathbf{u}) \hat{p}_{N,n}^{T}(\mathbf{u}) d\mathbf{u} d\mathbf{v}$$

$$= \lim_{N \to \infty} \sqrt{\frac{A_T A_R}{|\mathcal{S}|^2}} \int_{\mathcal{S}} \overline{p_m(\mathbf{v})} \int_{\mathcal{S}} H(\mathbf{v}, \mathbf{u}) p_n(\mathbf{u}) d\mathbf{u} d\mathbf{v} = \sqrt{\frac{A_T A_R}{|\mathcal{S}|^2}} v_n \delta_{mn}.$$

If we now define the sets $\{Q_N\}_{N=1}^{\infty}$ and $\{R_N\}_{N=1}^{\infty}$ from Equation (3.32) as $\{Q_N = \{f_{N,n}^T(\mathbf{u})\}_{n=1}^{K}\}_{N=1}^{\infty}$ and $\{R_N = \{g_{N,m}^R(\mathbf{u})\}_{m=1}^{K}\}_{N=1}^{\infty}$, respectively, then (3.32) can be rewritten as

$$\mathbf{y}_N = \mathbf{H}_N \mathbf{x}_N + \mathbf{n}_N, \tag{3.45}$$

where $\mathbf{y}_N = \{y_{N,1}, y_{N,2}, \ldots, y_{N,K}\}$ is the vector of instantaneous outputs from the receiver antennas $\{g_{N,m}^R(\mathbf{v})\}_{m=1}^{K}$, $\mathbf{x}_N = \{x_{N,1}, x_{N,2}, \ldots, x_{N,K}\}$ is the vector of instantaneous inputs to the transmitter antennas $\{f_{N,n}^T(\mathbf{u})\}_{n=1}^{K}$, $\mathbf{n}_N = \{n_{N,1}, n_{N,2}, \ldots, n_{N,K}\}$ is the vector of independent additive white Gaussian noise processes with power spectral density $N_0$ at the receiver antenna outputs, which are given by

$$n_{N,m} = \int_{\mathcal{A}_{N,m}^R} \bar{g}_{N,m}^R(\mathbf{v}) n(\mathbf{v}) d\mathbf{v}, \tag{3.46}$$

and the elements of the instantaneous $K \times K$ channel matrix $\mathbf{H}_N = (h_{m,n}^N)$ are given by

$$h_{m,n}^N = \int_{\mathcal{A}_{T,N}} \bar{g}_{N,m}^R(\mathbf{v}) \int_{\mathcal{A}_{R,n}} H(\mathbf{v},\mathbf{u}) f_{N,n}^T(\mathbf{u}) d\mathbf{u} d\mathbf{v} \xrightarrow[N \to \infty]{} \sqrt{\frac{A_T A_R}{|\mathcal{S}|^2}} v_n \delta_{mn}. \tag{3.47}$$

To derive the sequence of power contraints $E\{|x_{N,k}|^2\} = P_{N,k}$, for $k = 1, 2, \ldots, K$, corresponding to (3.33), we note that the lower bound in (3.20) can be rewritten as

$$\sum_{k=1}^{K} \log_2\left(\frac{|v_k|^2}{BN_0}\left[\frac{A_T A_R P}{K |\mathcal{S}|^2} + \frac{1}{K} \sum_{i=1}^{K} \frac{BN_0}{|v_i|^2}\right]\right)$$
$$= \sum_{k=1}^{K} \log_2\left(1 + \frac{A_T A_R |v_k|^2}{BN_0 |\mathcal{S}|^2}\left[\frac{P}{K} + \frac{|\mathcal{S}|^2}{KA_T A_R}\sum_{i=1}^{K}\frac{BN_0}{|v_i|^2} - \frac{|\mathcal{S}|^2 BN_0}{A_T A_R |v_k|^2}\right]\right), \tag{3.48}$$

and the quantities

$$\left[\frac{P}{K} + \frac{|\mathcal{S}|^2}{KA_T A_R}\sum_{i=1}^{K}\frac{BN_0}{|v_i|^2} - \frac{|\mathcal{S}|^2 BN_0}{A_T A_R |v_k|^2}\right], \quad k = 1, 2, \ldots, K, \tag{3.49}$$

are all guaranteed to be positive by (3.21) with

$$\sum_{k=1}^{K}\left[\frac{P}{K}+\frac{|S|^2}{KA_TA_R}\sum_{i=1}^{K}\frac{BN_0}{|v_i|^2}-\frac{|S|^2 BN_0}{A_TA_R|v_k|^2}\right]=P+\frac{|S|^2}{A_TA_R}\sum_{i=1}^{K}\frac{BN_0}{|v_i|^2}-\frac{|S|^2}{A_TA_R}\sum_{i=1}^{K}\frac{BN_0}{|v_i|^2}=P. \quad (3.50)$$

Hence, if we adopt the sequence of power constraints $\{P_N=\{P_{N,k}\}_{k=1}^{K}\}_{N=1}^{\infty}$ given by

$$E\{|x_{N,k}|^2\}=P_{N,k}=\frac{P}{K}+\frac{|S|^2}{KA_TA_R}\sum_{i=1}^{K}\frac{BN_0}{|v_i|^2}-\frac{|S|^2 BN_0}{A_TA_R|v_k|^2}, \quad k=1,2,\ldots,K;\; N=1,2,\ldots, \quad (3.51)$$

then the pair (3.45) and (3.51) give us a sequence of $K\times K$ observational models subject to a sequence of corresponding power constraints for which the sequence of channel matrices $\mathbf{H}_N$, $N=1,2,\ldots$, have singular values $\{v_{N,k}\}_{k=1}^{K}$ that satisfy

$$\lim_{N\to\infty}|v_{N,k}|^2=\frac{A_TA_R}{|S|^2}|v_k|^2, \quad k=1,2,\ldots,K.$$

If we let $\xi_{\{\mathbf{H}_N,P_N\}}$ represent the maximum achievable spectral efficiency for the model (3.45) with power constraints (3.51), then we clearly have

$$\begin{aligned}\lim_{N\to\infty}\xi_{\{\mathbf{H}_N,P_N\}}&=\lim_{N\to\infty}\sum_{k=1}^{K}\log_2\left(1+|v_{N,k}|^2\frac{P_{N,k}}{BN_0}\right)\\&=\sum_{k=1}^{K}\log_2\left(1+\frac{A_TA_R|v_k|^2}{BN_0|S|^2}\left[\frac{P}{K}+\frac{|S|^2}{KA_TA_R}\sum_{i=1}^{K}\frac{BN_0}{|v_i|^2}-\frac{|S|^2 BN_0}{A_TA_R|v_k|^2}\right]\right)\\&=\sum_{k=1}^{K}\log_2\left(\frac{|v_k|^2}{BN_0}\left[\frac{A_TA_RP}{K|S|^2}+\frac{1}{K}\sum_{i=1}^{K}\frac{BN_0}{|v_i|^2}\right]\right),\end{aligned} \quad (3.52)$$

as desired. This proves the theorem. ∎

**Corollary 1.** For $3.9215\approx(\varepsilon_0-1)<\gamma g\ll(\varepsilon_0-1)\pi^4(d/\lambda)^4$, we have

$$\xi_{\gamma g}\approx\sqrt{(\gamma g)/(\varepsilon_0-1)}\log_2(\varepsilon_0)\approx 1.1610\sqrt{\gamma g}. \quad (3.53)$$

**Proof.** Let $\varepsilon_0-1<\gamma g\ll(\varepsilon_0-1)\pi^4(d/\lambda)^4$ and $|S|^2/(\lambda^2 d^2)=\sqrt{(\gamma g)/(\varepsilon_0-1)}$. Then, $|S|\ll\pi d^2$, as required. Recall from Property P2 above that $|v_1|^2\approx|v_2|^2\approx\cdots\approx|v_M|^2\approx L<1$ and

$\sum_{n=\mathcal{M}+1}^{\infty} |v_n|^2 \approx 0$, where $\mathcal{M} = \left\lceil |\mathcal{S}|^2 / \lambda^2 d^2 \right\rceil$. Hence, in this case the lower bound in (3.20) is well approximated as

$$\sum_{k=1}^{K} \log_2\left( \frac{|v_k|^2}{BN_0} \left[ \frac{A_T A_R P}{K|\mathcal{S}|^2} + \frac{1}{K}\sum_{i=1}^{K} \frac{BN_0}{|v_i|^2} \right] \right) \approx \sum_{n=1}^{\mathcal{M}} \log_2\left( \frac{L}{BN_0} \left[ \frac{A_T A_R P}{\mathcal{M}|\mathcal{S}|^2} + \frac{1}{\mathcal{M}}\sum_{i=1}^{\mathcal{M}} \frac{BN_0}{L} \right] \right)$$

$$\approx \mathcal{M} \log_2\left( 1 + \frac{A_T A_R LP}{\mathcal{M}|\mathcal{S}|^2 BN_0} \right) \approx \sqrt{\frac{\gamma g}{\varepsilon_0 - 1}} \log_2\left( 1 + \frac{A_T A_R LP}{\mathcal{M}^2 \lambda^2 d^2 BN_0} \right) \quad (3.54)$$

$$\approx \sqrt{\frac{\gamma g}{\varepsilon_0 - 1}} \log_2\left( 1 + \frac{\gamma g}{\mathcal{M}^2} \right) \approx \sqrt{\frac{\gamma g}{\varepsilon_0 - 1}} \log_2(\varepsilon_0) \approx 1.1610 \sqrt{\gamma g}.$$

Hence, from (3.20) we have

$$\sum_{k=1}^{K} \log_2\left( \frac{|v_k|^2}{BN_0} \left[ \frac{A_T A_R P}{K|\mathcal{S}|^2} + \frac{1}{K}\sum_{i=1}^{K} \frac{BN_0}{|v_i|^2} \right] \right) \approx \sqrt{\frac{\gamma g}{\varepsilon_0 - 1}} \log_2(\varepsilon_0) \leq \xi_{\gamma g} \leq \sqrt{\frac{\gamma g}{\varepsilon_0 - 1}} \log_2(\varepsilon_0), \quad (3.55)$$

or $\xi_{\gamma g} \approx \sqrt{(\gamma g)/(\varepsilon_0 - 1)} \log_2(\varepsilon_0) \approx 1.1610 \sqrt{\gamma g}$, as claimed. ∎

## IV. DISCUSSION

There are several salient characteristics regarding Theorem 1, Corollary 1, and their proofs that are worth mentioning at this point. With regard to the validity of approximation (3.53) given in Corollary 1, note that the requirement that $\gamma g \ll (\varepsilon_0 - 1)\pi^4 (d/\lambda)^4$ in the statement of the Corollary is satisfied in any realistic long-range scenario. This implies that the constraint $|\mathcal{S}| \ll \pi d^2$ is not an active constraint on the capacity of the aperture-constrained, free-space AWGN channel in any realistic scenario.

With regard to the accuracy of approximation (3.53), we note that it does not seem likely that for $\gamma g > \varepsilon_0 - 1$, the maximum value of the lower bound in (3.20) over all values of $|\mathcal{S}| \geq \max(A_T, A_R)$ is equal to the upper bound in (3.20); however, asymptotically, the two are essentially the same. To see this, let the value of the lower bound in (3.20) be represented as a function of $|\mathcal{S}|$ in the interval $|\mathcal{S}| \geq \max(A_T, A_R)$ as

$$\beta(|\mathcal{S}|) = \sum_{k=1}^{K(|\mathcal{S}|)} \log_2\left(\frac{|v_k|^2}{BN_0}\left[\frac{A_T A_R P}{K(|\mathcal{S}|)\cdot|\mathcal{S}|^2} + \frac{1}{K(|\mathcal{S}|)}\sum_{i=1}^{K(|\mathcal{S}|)}\frac{BN_0}{|v_i|^2}\right]\right),$$

$$K(|\mathcal{S}|) = \max\left\{\kappa \in \mathbb{Z}^+ : \frac{P}{\kappa} + \frac{|\mathcal{S}|^2}{\kappa A_T A_R}\sum_{i=1}^{\kappa}\frac{BN_0}{|v_i|^2} > \frac{|\mathcal{S}|^2 BN_0}{A_T A_R |v_k|^2}, \forall k = 1, 2, \ldots, \kappa\right\}.$$

(4.1)

Examination of the behavior of the set $\{|v_n|^2\}_{n=1}^{\infty}$ given in [25] for large values of $|\mathcal{S}|$ clearly indicates that the value of $\beta(|\mathcal{S}|)$ is continuous in $|\mathcal{S}|$, that the maximum value of $\beta(|\mathcal{S}|)$ is attained in the interval $|\mathcal{S}| \geq \max(A_T, A_R)$, that this maximum occurs for $|\mathcal{S}|^2/(\lambda^2 d^2) \approx \sqrt{(\gamma g)/(\varepsilon_0 - 1)}$, and that

$$\lim_{\gamma g \to \infty} \frac{\max_{|\mathcal{S}| \geq \max(A_T, A_R)} \beta(|\mathcal{S}|)}{\sqrt{\frac{\gamma g}{\varepsilon_0 - 1}} \log_2(\varepsilon_0)} = 1. \qquad (4.2)$$

Hence, in the sense of Equation (4.2), we have $\xi_{\gamma g} \to \sqrt{(\gamma g)/(\varepsilon_0 - 1)} \log_2(\varepsilon_0)$, asymptotically as $\gamma g \to \infty$. This implies that the return on investment for achieving $\gamma g > 3.9215$ in a deep-space communication system (or on any long-range, free-space communication channel) is really much greater than is generally understood. That is, conventional link-budget calculations, which are essentially derived from Equation (1.2), assume that $\xi_{\gamma g} \approx \log_2(1 + \gamma g)$, which is really only valid as long as $\gamma g \leq (\varepsilon_0 - 1) \approx 3.9215$. For $\gamma g > 3.9215$, we really have $\xi_{\gamma g} \approx 1.1610\sqrt{\gamma g}$, which implies that much higher data rates can be achieved by raising the basic received SNR figure above the weak-signal threshold of $\gamma g = 3.9215$ than is generally realized in the design of such communication systems.

As far as the achievability of $\xi_{\gamma g}$ is concerned, the proof of Theorem 1 shows that for the weak-signal regime of $\gamma g \leq 3.9215$, the channel capacity of $\xi_{\gamma g} = \log_2(1 + \gamma g)$ can be approached asymptotically by transmitting a single information stream from a single transmitting antenna to a single receiving antenna in the usual manner. That is, the information is transmitted in a stream of complex symbols $\{x_i\}_{i=1}^{\infty}$ at the rate of $B$ symbols per second (sps) using a code with rate $R < \xi_{\gamma g}$ chosen optimally from the class of codes generated randomly and independently using the distribution $x \sim \mathcal{N}(0, P)$.

On the other hand, the proof also shows that above the weak-signal regime, any spectral efficiency $\xi(|\mathcal{S}|)$ satisfying

$$\log_2(1+\gamma g) < \xi(|\mathcal{S}|) < \max_{|\mathcal{S}| \geq \max(A_T, A_R)} \beta(|\mathcal{S}|), \qquad (4.3)$$

where $\beta(|\mathcal{S}|)$ is given by Equation (4.1), can be achieved only by transmitting multiple independent data streams. Furthermore, any such rate $\xi(|\mathcal{S}|)$ can itself be approached asymptotically utilizing a physically-realizable MIMO antenna array with antenna elements distributed over the discs $\mathcal{S}_T$ and $\mathcal{S}_R$ having areas $|\mathcal{S}|$ centered at the information source and sink, respectively, where the number and distribution of elements in each array is not specified but can be determined by approximating a finite number of the eigenfunctions of the operator $\mathcal{H}_\mathcal{S}$ with simple functions. In fact, the antenna elements may be distributed throughout the entire spherical volumes $\mathcal{V}_T$ and $\mathcal{V}_R$, and only the projections onto $\mathcal{S}_T$ and $\mathcal{S}_R$ are determined by the chosen simple function approximations, which need not be the same at both ends of the link. The number of independent data streams transmitted over the chosen MIMO antenna array is $K(|\mathcal{S}|)$, also given by Equation (4.1), and is equivalent to the number of eigenfunctions $\{p_k\}_{k=1}^{K(|\mathcal{S}|)}$ of $\mathcal{H}_\mathcal{S}$ that must be approximated, but that number is independent of which simple function approximations are chosen to represent $\{p_k\}_{k=1}^{K(|\mathcal{S}|)}$ at each end of the link. Note that the sample points of the simple function approximations for $\{p_k\}_{k=1}^{K(|\mathcal{S}|)}$ at the transmitter must all be identical, and the sample points of the simple function approximations for $\{p_k\}_{k=1}^{K(|\mathcal{S}|)}$ at the receiver must also be identical, but the number and locations of those sample points need not be the same at both ends of the link. The sample point locations at each end of the link correspond to the locations in space of the MIMO antenna elements at each end of the link, and the sampled values of $\{p_k\}_{k=1}^{K(|\mathcal{S}|)}$ at those points correspond to the (complex-valued) weighting that must applied to the symbols in each independent data stream before transmission and reception of that stream. The powers of the independent data streams are given by

$$P_k = \frac{P}{K(|\mathcal{S}|)} + \frac{|\mathcal{S}|^2}{K(|\mathcal{S}|) \cdot A_T A_R} \sum_{i=1}^{K(|\mathcal{S}|)} \frac{BN_0}{|v_i|^2} - \frac{|\mathcal{S}|^2 BN_0}{A_T A_R |v_k|^2}, \quad k=1,2,\ldots,K(|\mathcal{S}|), \tag{4.4}$$

The information is transmitted in $K(|\mathcal{S}|)$ streams of complex symbols $\{x_{k,i}\}_{i=1}^{\infty}$, $k=1,2,\ldots,K(|\mathcal{S}|)$, each of which is transmitted at rate $B$ sps using individual codes of rates

$$R_k < \sum_{k=1}^{K(|\mathcal{S}|)} \log_2 \left( \frac{|v_k|^2}{BN_0} \left[ \frac{A_T A_R P}{K(|\mathcal{S}|) \cdot |\mathcal{S}|^2} + \frac{1}{K(|\mathcal{S}|)} \sum_{i=1}^{K(|\mathcal{S}|)} \frac{BN_0}{|v_i|^2} \right] \right), \tag{4.5}$$

chosen optimally from the class of codes generated randomly and independently using the distribution $x_k \sim \mathcal{N}(0, P_k)$ for $k=1,2,\ldots,K(|\mathcal{S}|)$.

Note that the question of how many individual antenna elements must be used to construct a set of suitable distributed antennas is not addressed in the proof of Theorem 1. It is clear from the proof of the theorem that all $K(|\mathcal{S}|)$ data streams can be transmitted from the same collection of distributed nodes, which together comprise $K(|\mathcal{S}|)$ different distributed antennas by varying the phase and amplitude of radiation across the fixed array of distributed elements. However, the number of nodes needed to achieve near-optimal performance is not addressed in the proof of the theorem. To determine the required number $N$ of elements in the distributed antenna array, it becomes necessary to study the smoothness characteristics of the two-dimensional prolate spheroidal wave functions, which is beyond the scope of this work; however, based on the Nyquist spatial sampling rate associated with the space-bandwidth characteristics of those functions, one can conjecture that the total number of required elements is approximately $N = O(|\mathcal{S}|^2 / \lambda^2 d^2)$. Using the same logic, one can also conjecture that as long as the $N$ nodes are fairly uniformly distributed throughout $\mathcal{S}$ at both ends of the link, the performance should be reasonably stable. That is, the overall achievable spectral efficiency should not be very sensitive to either the number of antenna nodes or small variations in the relative position of the nodes as long as the node density is above $|\mathcal{S}|/\lambda^2 d^2$ nodes per square meter and the distribution is roughly uniform. Further insight into this issue may be available in a recent paper on approximation of essentially space- and band-limited functions in two-dimensions [27].

# V. Conclusion

In this paper, we have studied the capacity of the long-range, free-space AWGN communication channel when available spatial diversity is incorporated into the channel model itself. In particular, the capacity of the channel has been studied for the scenario conceptually encountered in long-range deep-space communication applications. That is, information is assumed to be transmitted over a band-limited radio channel in free space over a very large distance from an information source (the transmitter) to an information sink (the receiver) subject to constraints on both the total transmitter power and the total effective aperture areas of the transmitter and receiver antennas; however, the antennas at both the transmitter and the receiver may be distributed in an arbitrary fashion over spherical regions of space at both ends of the communication link subject only to the constraint that the radii of both regions is much smaller than the distance between them. The following results have been presented.

1. We derived upper and lower bounds on the maximum achievable spectral efficiency for this channel and showed that, for values of the received SNR below a weak-signal threshold, these bounds are equal and equivalent to the well-known SISO result given by Equation (1.2).

2. For values above the weak-signal received SNR threshold, we also derived a simple approximation for the maximum achievable spectral efficiency, which shows that the capacity for the long-range free-space channel actually strictly exceeds the result given by Equation (1.2) and grows asymptotically as a function of the square root of the received SNR rather than logarithmically as in Equation (1.2).

3. We showed that for values above the weak-signal threshold, the lower bound on the maximum achievable spectral efficiency can be approached arbitrarily closely using physically realizable distributed MIMO antenna arrays satisfying the antenna aperture constraints at each end of the link. This may have significance for the design of future high-data-rate deep-space communication systems.

The implications of these results for the practical implementation of nearly optimal long-range, free-space communication systems has been discussed briefly, and further work in that area is anticipated in future studies.

## APPENDIX A – PROLATE SPHEROIDAL WAVE FUNCTIONS

Let $\mathcal{D}$ represent the unit circle in $\mathbb{R}^2$, and recall that $\mathcal{S} = R\mathcal{D}$. The eigenvalues and eigenfunctions for Problem (3.4) are given by

$$v_n = \sqrt{\frac{L}{\lambda^2 d^2}} R^2 \alpha_n,$$
$$p_n(\mathbf{u}) = \psi_n(\mathbf{u}/R), \quad \mathbf{u} \in \mathcal{S}, \tag{A.1}$$

where $\{\alpha_n\}_{n=1}^{\infty}$ and $\{\psi_n(\mathbf{u})\}_{n=1}^{\infty}$ are the eigenvalues and eigenfunctions of the integral equation

$$\alpha_n \psi_n(\mathbf{v}) = \iint_{\mathcal{D}} e^{i \frac{2\pi R^2}{\lambda d} \langle \mathbf{v}, \mathbf{u} \rangle} \psi_n(\mathbf{u}) d\mathbf{u}, \quad \mathbf{v} \in \mathcal{D}. \tag{A.2}$$

The solutions to this integral equation have been widely studied [24, 25] and are given by the *prolate spheroidal wave functions*

$$\psi_{|N|,m}(r,\theta) = R_{|N|,m}(r) e^{i \frac{2\pi}{\lambda d} N\theta}, \quad N = 0, \pm 1, \pm 2, \ldots, \ m = 0, 1, 2, \ldots, \tag{A.3}$$

(now doubly indexed and given in terms of the polar coordinates $\mathbf{u} = (r,\theta)$) and the associated eigenvalues

$$\alpha_{N,m} = 2\pi i^N \beta_{|N|,m}, \tag{A.4}$$

where

$$\beta_{|N|,m} R_{|N|,m}(r) = \int_0^1 J_{|N|}\left(\frac{2\pi R^2}{\lambda d} rr'\right) R_{|N|,m}(r') r' \, dr', \quad 0 \leq r \leq 1, \tag{A.5}$$

and $\{J_N(r)\}_{N=0}^{\infty}$ are the Bessel functions of the first kind [26].

## APPENDIX B – PROOF OF EQUATION (3.24)

Let $P_n = Q_n^2$ for all $n = 1, 2, \ldots$, and let

$$J(\mathbf{Q}, \lambda) = \sum_{n=1}^{\infty} \log_2\left(1 + |\eta_n|^2 \frac{P_n}{BN_0}\right) + \rho\left(P - \sum_{n=1}^{\infty} Q_n^2\right), \tag{B.1}$$

where $\mathbf{Q} = \{Q_n\}_{n=1}^{\infty}$ and $\rho$ represents a Lagrange multiplier for the constraint $\sum_{n=1}^{\infty} Q_n^2 = P$. Recall that $|\eta_n| \geq |\eta_{n+1}| > 0$ for all $n = 1, 2, \ldots$. Differentiating wrt $Q_n$ gives

$$\frac{\partial J}{\partial Q_n} = (\log_2 e) \left[ \left( \frac{2Q_n |\eta_n|^2}{BN_0} \right) \bigg/ \left( 1 + \frac{Q_n^2 |\eta_n|^2}{BN_0} \right) \right] - 2\rho Q_n, \tag{B.2}$$

so for any stationary point, we must have

$$\begin{aligned} 0 &= (\log_2 e) \left[ \left( \frac{2Q_n |\eta_n|^2}{BN_0} \right) \bigg/ \left( 1 + \frac{Q_n^2 |\eta_n|^2}{BN_0} \right) \right] - 2\rho Q_n \\ &= 2Q_n (\log_2 e) \left[ \frac{|\eta_n|^2}{BN_0 + Q_n^2 |\eta_n|^2} - \rho' \right], \quad \forall n = 1, 2, \ldots, \end{aligned} \tag{B.3}$$

where $\rho' = \rho / (\log_2 e)$. Hence, either $Q_n = 0$ or

$$\rho' = \frac{|\eta_n|^2}{BN_0 + Q_n^2 |\eta_n|^2} \Leftrightarrow Q_n^2 = \frac{1}{\rho'} - \frac{BN_0}{|\eta_n|^2}. \tag{B.4}$$

It follows that

$$Q_n^2 = \left[ \frac{1}{\rho'} - \frac{BN_0}{|\eta_n|^2} \right]^+, \quad \forall n = 1, 2, \ldots, \tag{B.5}$$

whence applying the power constraint gives

$$P = \sum_{n=1}^{\infty} \left[ \frac{1}{\rho'} - \frac{BN_0}{|\eta_n|^2} \right]^+ = \sum_{n=1}^{K} \left[ \frac{1}{\rho'} - \frac{BN_0}{|\eta_n|^2} \right], \tag{B.5}$$

where $K$ is the greatest integer such that

$$\frac{1}{\rho'} > \frac{BN_0}{|\eta_n|^2}, \quad \forall n = 1, 2, \ldots, K. \tag{B.6}$$

Solving (B.5) for $\rho'$ establishes (3.24), as desired. ∎

REFERENCES

[1] R. J. Barton, "Distributed MIMO Communication Using Small Satellite Constellations," *Proceedings of the 2014 IEEE International Conference on Wireless for Space and Extreme Environments (WiSEE 2014)*, pp. 1-7, October 2014.

[2] E. Teletar, "Capacity of Multi-antenna Gaussian Channels," *European Transactions on Telecommunications*, vol. 10, issue, 6, pp. 585-595, Dec. 1999.

[3] G. J. Foschini, G. J. and M. J. Gans, "On Limits of Wireless Communications in a Fading Environment when Using Multiple Antennas," *Wireless Personnel Communications*, vol. 6, pp. 311-335, 1998.

[4] T. L. Marzetta and B. M. Hochwald, "Capacity of a Mobile Multiple-Antenna Communication Link in Rayleigh Flat Fading," *IEEE Transactions on Information Theory*, vol. 45, no. 1, pp. 139-157, Jan. 1999.

[5] D.-S. Shiu, G. J. Foschini, M. J. Gans, and J. M. Kahn, "Fading Correlation and Its Effect on the Capacity of Multielement Antenna Systems," *IEEE Transactions on Communications*, vol. 48, no. 3, pp. 502-513, March 2000.

[6] C.-N. Chuah, D. N. C. Tse, J. M. Kahn, and R. A. Valenzuela, "Capacity Scaling in MIMO Wireless Systems Under Correlated Fading," *IEEE Transactions on Information Theory*, vol. 48, no. 2, pp. 637-650, March 2002.

[7] D. A. B. Miller, "Communicating with Waves Between Volumes: Evaluating Orthogonal Spatial Channels and Limits on Coupling Strengths," *Applied Optics*, vol. 39, no. 11, pp. 1681-1699, April 10, 2000.

[8] A. M. Sayeed, "Deconstructing Multiantenna Fading Channels," *IEEE Transactions on Signal Processing*, vol. 50, no. 10, pp. 2563-2579, Oct. 2002.

[9] T. S. Pollock, T. D. Abhayapala, and R. A. Kennedy, "Antenna Saturation Effects on MIMO Capacity," *Proceedings of the International Conference on Communications (ICC)*, vol. 3, pp. 2301-2305, May 2003.

[10] T. D. Abhayapala, T. S. Pollock, and R. A. Kennedy, "Characterization of 3D Spatial Wireless Channels,' *IEEE 58th Vehicular Technology Conference*, vol. 1, pp. 123-127, Oct. 6-9, 2003.

[11] A. S. Y. Poon, R. W. Brodersen, and D. N. C. Tse, "Degrees of Freedom in Multiple-Antenna Channels: A Signal Space Approach, *IEEE Transactions on Information Theory*, vol. 51, no. 2, pp. 523-536, Feb. 2005.

[12] L. Hanlen and M. Fu, "Wireless Communication Systems With Spatial Diversity: A Volumetric Model, *IEEE Transactions on Wireless Communications*, vol. 5, no. 1, pp. 133-142, Jan. 2006.

[13] P. Gupta and P. R. Kumar, "The Capacity of Wireless Networks," *IEEE Transactions on Information Theory*, vol. 42, no. 2, pp. 388-404, 2000.

[14] L. -L. Xie and P. R. Kumar, "A Network Information Theory for Wireless Communications: Scaling Laws and Optimal Operation," *IEEE Transactions on Information Theory*, vol. 50, no. 5, pp. 748-767, 2004.

[15] A. Ozgur, O. Leveque, D. Tse, "Hierarchical Cooperation Achieves Optimal Capacity Scaling in Ad-Hoc Networks," *IEEE Transactions on Information Theory*, vol. 53, no. 10, pp. 3549-3572, 2007.

[16] M. Franceschetti, M. D. Migliore, P. Minero, "The Capacity of Wireless Networks: Information-theoretic and Physical Limits," *IEEE Transactions on Information Theory*, vol. 55, no. 8, pp. 3413–3424, August 2009.

[17] A. Ozgur, O. Leveque, D. Tse, "Spatial Degrees of Freedom of Large Distributed MIMO Systems and Wireless Ad-Hoc Networks," *IEEE Journal on Selected Areas in Communications*," vol. 31, no. 2, February 2013.

[18] P.-D. Arapoglou, K. Liolis, M. Bertinelli, A. Panagopoulos, P. Cottis, and R. De Gaudenzi, "MIMO over Satellite: A Review," *IEEE Communications Surveys and Tutorials*, vol. 13, no. 1, pp. 27-51, First Quarter 2011.

[19] T. M. Cover and J. A. Thomas, *Elements of Information Theory*, 2nd Edition, John Wiley & Sons, New York, 2006.

[20] J. G. Proakis, Digital Communications, 4th Edition, McGraw-Hill, New York, 2001.

[21] E. Bigleiri, R. Calderbank, A. Constantinides, A. Goldsmith, A. Paulraj, H. V. Poor, *MIMO Wireless Communications*, Cambridge University Press, Cambridge, 2007.

[22] A. W. Naylor and G. R. Sell, *Linear Operator Theory in Engineering and Science*, 2nd Edition, Springer-Verlag, New York, 1982.

[23] J. Bell, "The Singular Value Decomposition of Compact Operators on Hilbert spaces," http://individual.utoronto.ca/jordanbell/notes/SVD.pdf, 2014, unpublished.

[24] D. Slepian, "Prolate Spheroidal Wave Functions, Fourier Analysis and Uncertainty-IV: Extensions To Many Dimensions; Generalized Prolate Spheroidal Functions," *Bell System Technical Journal*, vol. 43, pp. 3009-3058, 1964.

[25] H. J. Landau, "On Szegö's Eingenvalue Distribution Theorem and Non-Hermitian Kernels, *Journal d'Analyse Mathématique*, vol. 28, Issue 1, pp. 335-357, December 1975.

[26] M. Abramovitz and I. A. Stegun, Editors, *Handbook of Mathematical Functions: with Formulas, Graphs, and Mathematical Tables*, Dover, New York, 1965.

[27] B. Landa and Y. Shkolniski, "Approximation scheme for essentially bandlimited and space-concentrated functions on a disk," *Applied and Computational Harmonic Analysis*, 2016, http://www.sciencedirect.com/science/article/pii/S1063520316000075.